\documentclass{elsart}

\usepackage{graphicx}

\usepackage{amssymb}

\begin{document}
\begin{frontmatter}

\title{ Large-System Phase-Space Dimensionality Loss
        in Stationary Heat Flows }

\author[a]{Harald A. Posch},
\ead{posch@ls.exp.univie.ac.at}
\author[b]{William G. Hoover}
\ead{hoover3@llnl.gov}
\address[a]{Institute for Experimental Physics, University of Vienna,
Boltzmanngasse 5, A-1090 Vienna, Austria}
\address[b]{Department of Applied Science, University of California at
        Davis/Livermore and Methods Development Group, Lawrence Livermore
        National Laboratory, Livermore, California 94551-7808, USA}
\date{\today}

\begin{abstract}

	Thermostated tethered harmonic lattices provide good illustrations
of the phase-space dimensionality loss $\Delta D$ which occurs
in the strange-attractor distributions characterizing stationary
nonequilibrium flows.  We use time-reversible nonequilibrium molecular
dynamics, with two Nos\'e-Hoover thermostats, one hot and one cold,
to study a family of square heat-conducting systems.  We find a
phase-space dimensionality loss which can exceed the dimensionality 
associated with the two driving Nos\'e-Hoover thermostats by as much as
a factor of four.  We also estimate the dimensionality loss
$\Delta D_{\mathcal{H}}$ in the purely Hamiltonian part of phase space.  By
measuring the {\em projection} of the total dimensionality loss there we
show that nearly all of the loss occurs in the Hamiltonian part.  Thus
this loss, which characterizes the extreme rarity of nonequilibrium
states, persists in the large-system thermodynamic limit.

\end{abstract}

\end{frontmatter}


\section{Introduction}

	Aoki and Kusnezov stressed the usefulness of the ``$\phi ^4$"
model in studying the size-dependence of classical heat conductivity
simulations \cite{b1}.  They used this model to show that the entropy
production associated with steady heat flow becomes extensive,
($\propto N$), as the system size $N$ increases.  When such heat-flow
simulations are driven with two
time-reversible Nos\'e-Hoover thermostats \cite{b2,b3,b4,b5,b6,b7}
imposing two temperatures, ``hot" and ``cold", the steady-state
phase-space distribution function becomes a multifractal
object \cite{b8,b9,b10}.  The information dimension of this multifractal
object lies below the corresponding
equilibrium one by an extensive ($\propto N$) ``dimensionality
loss" $\Delta D$.  The dimensionality loss, like the closely-related
thermodynamic entropy production $\dot S$, is approximately
quadratic in the deviation from equilibrium brought about by the
temperature gradient $\nabla T$:
$$ \Delta D \propto N(\nabla \ln T)^2 .$$
This quadratic dependence is the usual prediction of linear
transport theory.  Its evaluation quantifies the rarity of nonequilibrium
stationary states.

	About ten years ago \cite{b11} we studied phase-space dimensionality
loss in a family of two-dimensional shear flows.  Those simulations were
limited to relatively small systems.  In that shear-flow work we 
were able to find systems for which the dimensionality loss barely, but
significantly, exceeded the total dimensionality associated with the
coordinates, momenta, and friction coefficients $\{ q,p,\zeta \}$ of the
thermostated boundaries.

	At that time we stated that our data definitely
showed that nonequilibrium stationary states can exhibit a reduced
dimensionality, relative to equilibrium, even when projected into the
subspace occupied by purely-Hamiltonian degrees of freedom.  This
interpretation was controversial \cite{b12,b13}.  Many researchers had difficulty in
accepting that nonequilibrium stationary states are typically associated with
reduced dimensionality.  In order to confirm that our interpretation was
correct, we  recently studied dimensionality reduction for stationary flows
using the (Hamiltonian) $\phi ^4$ model.  Because heat transfer using this
model requires only the simplest of boundaries---a single hot degree of
freedom and a single cold one, in the simplest one-dimensional case---it
turned out to be easy to get relatively large dimensionality losses \cite{b2}.

\begin{figure}
\centering{\includegraphics[width=8cm,clip=]{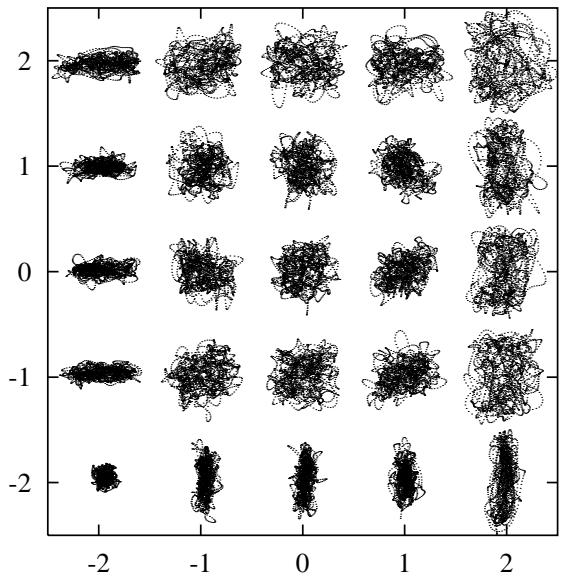}}
\caption{
Geometry of a 25-particle two-dimensional steady-state system with a
full-phase space dimensionality reduction $\Delta D = 21.6$. The upper
righthand particle is ``hot" and the lower lefthand one is ``cold".
The remaining 23 particles obey conventional Hamiltonian mechanics.  The
trajectories shown were generated with hot and cold
temperatures $T_H=0.009$ and $T_C=0.001$ imposed by two Nos\'e-Hoover
thermostats with relaxation times of $\tau = 5$.
}
\label{fig-1}
\end{figure}

	A simple but convincing two-dimensional
case involves the heat transfer from a single hot particle (with five
phase-space coordinates $\{ x_H,y_H,p_{xH},p_{yH},\zeta _H \} $) to a
single cold one (with its own five coordinates) through a medium of
$N-2=23$ Hamiltonian particles (with 92 more phase-space coordinates, for a
total of $4N+2=102$ phase-space coordinates).  Typical $(x,y)$ particle
trajectories in such a 25-particle system are shown in Fig. \ref{fig-1}.  
In the
case of $N=16$ particles discussed in Ref. 2 the 
phase-space strange attractor turned out to have a dimensionality 
loss of 12.5, relative to the equilibrium distribution, exceeding the
number of thermostated phase-space coordinates (5 hot and 5 cold) by 2.5.

	The present work has two goals, (i) characterizing the overall
loss of phase-space dimensionality $\Delta D$, and (ii) estimating the
corresponding loss $\Delta D_{\mathcal{H}}$ in the Hamiltonian unthermostated
part of phase space.
This two-part study requires the numerical evaluation of the instantaneous
Lyapunov exponents $\{ \lambda (t)\} $, as well as their associated
vectors $\{ \delta \} $.  The required
techniques are sketched in Sec. II.  In Sec. III we describe the $\phi ^4$
model.  In Sec. IV we undertake a
systematic study of the {\em number-dependence} of the phase-space
dimensionality loss $\Delta D$.  We consider a family of systems of the
type illustrated in Fig. 1.  The numerical work
described in Sec. IV shows convincingly a $\sqrt {1/N}$ deviation from the
large-system limit, with an extrapolated dimensionality loss of as much as
four times (40) the number of dimensions (10) associated with the thermal
driving mechanism.  Next, in Sec. V, we study the projection of this
dimensionality loss into the purely Hamiltonian subspace.  We find that
nearly all of the loss occurs in that subspace despite the purely
Hamiltonian form of the equations of motion there.  We discuss the projection
technique and the importance of rapid {\em rotation} in phase space \cite{b14}
to an understanding of the strange-attractor's dimensionality reduction.  The
conclusions which follow from this work take up the final section.

\section{Lyapunov Exponent Calculations}

	The usual ``Lyapunov exponents", $\{ \lambda _j\} $, are time averages,
over a sufficiently long time for convergence, of ``local" (instantaneous)
exponents $\{ \lambda _j(t)\} $,
$$ \lambda _j \equiv \langle \lambda _j(t) \rangle \ .$$
Here, and in what follows, we use angular brackets $\langle \dots \rangle $
to indicate long-time averages.
Each of the instantaneous exponents has
associated with it an ``offset" vector $\delta _j$ in phase space which
describes the direction in which the growth or decay of phase-space
separation is measured.  If the offset vectors are infinitesimal in length,
as in the present work, they can be replaced by a parallel set of unit 
vectors in ``tangent space".

	The time evolution of the offset vector directions is governed by a
continuous Gram-Schmidt orthonormalization which forces the
vectors to remain mutually perpendicular and to evolve at fixed length.  The
two types of constraints, orthogonality and fixed length, can be imposed by
a triangular array of Lagrange multipliers, $\Lambda _{ij}$, where 
$1\le j\le i\le n$ in an $n$-dimensional phase space \cite{b15}.
When the usual Cartesian coordinates are used the orthonormal offset vetors
rotate rapidly \cite{b14}.

	The usual Lyapunov exponents $\{ \lambda _{j}\} $ are the
long-time-averaged values of the diagonal Lagrange multipliers
$\{ \Lambda _{jj}\} $ required to enforce the orthonormalization constraints:
$$ \lambda _j = \langle \lambda _j(t) \rangle \equiv
\langle \Lambda _{jj} \rangle \ .$$
At any instant of time the instantaneous Lyapunov exponents $\lambda (t)$
represent the orthogonal measurements of the $n$ growth (or decay) rates
determined by the dynamical matrix $D$.  $D$, an $n\times n$ matrix,
is itself a phase function.  The $i$th row of $D$ is made up of the $n$
derivatives of the $i$th equation of motion with respect to the $n$ phase
variables.

	The eigenvalues and eigenvectors of the dynamical matrix are
relatively complicated \cite{b16}.  The eigenvalues are mainly rapidly
varying complex-conjugate pairs, with the imaginary parts frequently
vanishing, at phase-space singularities corresponding to parallel
eigenvectors.  Despite this complexity the orthonormal basis provided by
the offset vectors makes it possible to measure smooth and well-defined
growth rates.  The time reversibility of the equations of motion guarantees
related reversibility properties for $D$ and its eigenvectors and
eigenvalues.  Nevertheless, the past-based Lyapunov vectors governed by
$D$ show a time-symmetry breaking intimately related to the second law of
thermodynamics \cite{b3,b8,b9,b11,b14,b16}.

\begin{figure}
\centering{\includegraphics[width=10cm,clip=]{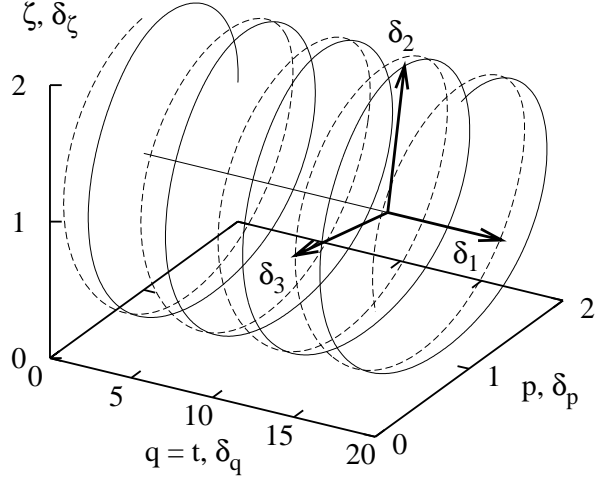}}
\centering{\includegraphics[width=8cm,clip=]{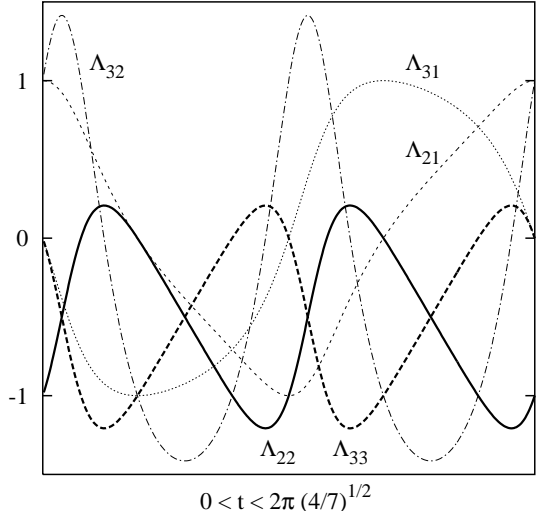}}
\caption{
Bottom: Time dependence of the five nonvanishing Lagrange multipliers for a
field-driven thermostated particle in one dimension,
as discussed in Sec. II.  The time dependence of the vectors 
$\{ \delta _2,\delta _3\} $ for this same problem is shown at the top.
The period of the oscillation is $2\pi \sqrt{4/7}$ though the phase
has no significance in the stationary state.
}
\label{fig-2}
\end{figure}

	Consider a simple textbook \cite{b3}(Sec. 5.4) illustration of these
ideas, the motion of a driven thermostated particle in one dimension.  
The motion takes place in the three-dimensional $(q,p,\zeta )$ phase space
with three Lyapunov exponents:
$$
\begin{array}{ccccc}
\lambda _1 &=& \langle \lambda _1(t)\rangle &=& \langle \Lambda _{11} \rangle 
                     \nonumber \\ 
\lambda _2 &=& \langle \lambda _2(t)\rangle &=& \langle \Lambda _{22} \rangle 
                     \nonumber \\ 
\lambda _3 &=& \langle \lambda _3(t)\rangle &=& \langle \Lambda _{33} \rangle 
                      \; .  \nonumber  
\end{array}
$$
The equations of motion (with unit mass, force constant, and relaxation time)
are
$$\dot q = p\ ;\ \dot p = +1 - \zeta p\ ;\ \dot \zeta = p^2 - 1\ ,$$
for which the long-time solution is an attractor:
$$\{ q(t),p(t),\zeta (t)\} \longrightarrow \{ t,+1,+1\} \ .$$
The matrix of equation-of-motion derivatives $D$ is
$$D =  \left(
\begin{array}{ccc}
                                      0  & +1     &  0    \\
                                      0  & -\zeta & -p    \\
                                      0  & 2p     &  0
\end{array}
\right) \ .$$
On the attractor the momentum $p$ and friction coefficient $\zeta $ can be
replaced by their limiting values, (+1,+1).
The three tangent-space $\delta $ vectors follow the ordinary differential
equations:
$$
\begin{array}{cccccccc}
\dot \delta _1 =& D\cdot \delta _1& - & \Lambda _{11}\delta _1 &&&&\nonumber \\
\dot \delta _2 =& D\cdot \delta _2& - & \Lambda _{21}\delta _1 
                                  & - & \Lambda _{22}\delta _2 &&\nonumber \\
\dot \delta _3 =& D\cdot \delta _3& - & \Lambda _{31}\delta _1
                                  & - & \Lambda _{32}\delta _2
                                  & - & \Lambda _{33}\delta _3 \;, \nonumber
\end{array}
$$
where the six Lagrange multipliers follow easily from the time derivatives
of the six orthonormality conditions:
$$ \delta _1^2 = \delta_2^2 = \delta_3^2 \equiv 1\ ;$$
$$ \delta _1\cdot \delta _2 =
   \delta _2\cdot \delta _3 =
   \delta _3\cdot \delta
_1 \equiv 0\ .$$
The steady-state time variation of the Lagrange multipliers, and the three
$\delta $ vectors are
shown in Fig. \ref{fig-2}.  The three Lyapunov exponents, like their three
instantaneous values $\{ \lambda _j(t) = \Lambda _{jj} \} $, sum to $-1$:
$$\frac{\partial \dot q    }{\partial q     } +
  \frac{\partial \dot p    }{\partial p     } +
  \frac{\partial \dot \zeta}{\partial \zeta } \equiv
  \lambda _1(t) + \lambda _2(t) + \lambda _3(t) = -1\ .$$
Note that the largest (time-averaged) Lyapunov exponent is 0 in this case,
corresponding to a phase-space displacement in the direction of the motion
$\delta _1 = (1,0,0)$.  The remaining five Lagrange multipliers, as well as
the vectors $\delta _2$ and $\delta _3$, oscillate
periodically, with a period of $2\pi \sqrt{4/7}$.

	An analogous computation for an $n$-dimensional set of vectors involves
computational work of order $n^3$.  Nonequilibrium steady states generate
chaotic attractors rather than the simple fixed point of the damped oscillator
example.  Nevertheless, the basic steps are the same: (i) propagating a
phase-space reference trajectory; (ii) simultaneously propagating offset
vectors in the neighborhood of the reference, with Lagrange multipliers and/or
Gram-Schmidt orthonormalization imposing orthonormality; and (iii) averaging
the diagonal multipliers to find the Lyapunov spectrum.  Let us now consider
the many-body model for which we carry out such investigations.

\section{$\phi ^4$ Heat Flow in Two Dimensions}

	In the ``$\phi ^4$'' model we use here \cite{b1,b2} we choose a square
nearest-neighbor harmonic lattice with a quadratic Hooke's Law pair potential
for neighboring particles $i$ and $j$,
$$ \phi (r) = \textstyle{\frac{\kappa _2}{2}}(r-d)^2\ ;
\ r = |r_i-r_j| > 0 \ .$$
In addition, each particle is tethered to its lattice site with a {\em quartic}
potential, $\textstyle{\frac{\kappa _4}{4}}\delta r^4$.  The quartic tethers
have two nice consequences.  First, they provide ``external forces", and so
prevent momentum conservation and ballistic energy transport.  Second, they
can provide {\em chaos}, so that the dynamics can have one or more
positive Lyapunov exponents.  The two consequences together can give
Fourier heat conduction, even in one or two space dimensions.  The heat
conductivity for the $\phi ^4$ model remains finite in the large-system
limit, unlike many models, for which the conductivity vanishes or diverges
in the large-system limit.

	Here we simulate nonequilibrium heat-conducting stationary states by
imposing thermostating forces on two of the $N$ particles.  The thermostating
forces are Nos\'e-Hoover \cite{b3,b4} feedback forces $\{ - \zeta p \} $,
linear in both the time-reversible friction coefficients $\{ \zeta \} $ and
the momenta $\{ p\}  = \{ (p_x,p_y)\} $.  The thermostated equations of
motion for the hot and cold particles (one of each) are:
$$
\begin{array}{ccclccclc}
\{&  m\dot x  &=& p_x  &;& m\dot y &=& p_y &\} \nonumber \\
\{&  \dot p_x &=& F_x - \zeta p_x &;& 
       \dot p_y &=& F_y - \zeta p_y  &\} \nonumber
\end{array}
$$
$$\dot \zeta_{(H\ {\rm or}\ C)} =
[(p^2/2mkT_{(H\ {\rm or}\ C)})-1]/\tau^2 \ ,$$
where $\tau $ is the characteristic response time of the thermostat
forces, $\{ -\zeta p\} $.  The full phase space describing this $N$-body
two-dimensional system has $4N+2$ dimensions, with the extra two corresponding
to the hot and cold friction coefficients $\zeta _H$ and $\zeta _C$.

	For convenience in numerical work we choose the particle mass $m$,
the spring constants $\kappa _2$ and $\kappa _4$, and the nearest-neighbor
lattice spacing $d$ all equal to unity.  The remaining parameters to set
are the hot and cold temperatures, which we arbitrarily choose equal to
the values from Ref. 2,
$$kT_H = 0.009\ ;\ kT_C = 0.001\ ,$$
and the thermostat relaxation times.  We vary these times in the numerical
work, but with the simplifying restriction that $\tau _H$ and $\tau _C$
have a common value, which we denote as $\tau $.
We use the classic fourth-order Runge-Kutta integrator throughout, with
a timestep $dt = 0.001\ {\rm or}\ 0.002$.  To avoid numerical errors we have generated and
compared results from two fully-independent simulation codes, one written
in Vienna and the other written in Livermore.
 
\section{Numerical Results---$\Delta D$}

	The numerical evaluation of the dimensionality loss $\Delta D$
is based on the connection between the Lyapunov
spectrum and the dimensionality of the phase-space strange attractor.  The
Lyapunov exponents give the time-averaged relative growth and decay rates
of the principal axes of a comoving infinitesimal phase-space hypersphere
(or ``extension in phase").  Kaplan and Yorke conjectured, evidently
correctly \cite{b17}, that a partial sum of these exponents (beginning with
the largest one and
proceeding toward the most negative one) changes from positive to negative
when the (linearly-interpolated) number of terms in the sum is equal to the
dimensionality of the phase-space strange attractor.  This conjecture is
``almost obvious".  It is evident that the (hyper)volume of any phase-space
object with a positive sum of time-averaged Lyapunov exponents must diverge.
Likewise a negative sum of time-averaged exponents indicates a vanishing
hypervolume at long times.  Any stationary process must generate an attractor
which neither vanishes nor diverges.

	The main difficulty in computing Kaplan-Yorke information
dimensions is the unfavorable time-dependence associated with constraining
the phase-space offset vectors to remain perpendicular to one another.  With
Gram-Schmidt orthonormalization in an $n$-dimensional phase space $n$ vectors,
with $n$ components each, must all be propagated in time
for sufficiently long that the time-averaged growth rates have converged.
The computational work in orthogonalizing $n(n-1)/2$ pairs of $n$-dimensional
vectors varies as the {\em cube} of the number of particles, so that present
computer speeds and processor numbers allow us to follow no more than
a few hundred particles.  In the present work we consider the simplest
possible square
systems of from $4 (2\times 2)$ to $144 (12\times 12)$ particles.  The
relaxation time $\tau $ is a free parameter.  We choose it in the range 
$1\dots 8$.  Results become insensitive to $\tau $ once $\tau $ exceeds 6,
which is comparable to the inverse Debye frequency of the lattice.
Representative results for the largest Lyapunov exponent $\lambda _1$ and the
dimensionality loss from the Kaplan-Yorke conjecture, $\Delta D$, are given
in Table \ref{table1}.
\begin{table}
\caption{
Representative data for the total phase-space dimensionality loss $\Delta D$
and the Hamiltonian projection $\Delta D_{\mathcal{H}}$ as a function of
system size and the Nos\'e-Hoover relaxation time $\tau $.
The largest Lyapunov exponent, $\lambda _1$ is also tabulated.
The two boundary temperatures (imposed by a single hot and a
single cold particle) are 0.009 and 0.001 in all cases. The estimated 
error is $\pm 0.4$ for $\Delta D$  and  $\Delta D_{\mathcal{H}}$, and    
$\pm 0.0005$ for $\lambda _1$.
}
\label{table1}

\begin{center}
\begin{tabular}{ccccc|ccccc}
$ \sqrt{N} $ & $ \tau $ & $ \Delta D $ & $ 
    \Delta D _{\mathcal{H}}$ & $ \lambda _1   $ &
$ \sqrt{N} $ & $ \tau $ & $ \Delta D $ & $ 
    \Delta D _{\mathcal{H}}$ & $ \lambda _1   $  \\ \hline

$ 4 $ & $ 1 $ & $ 10.3 $ & $ 10.6 $ & $ 0.0633  $     & 
$ 4 $ & $ 4 $ & $ 17.6 $ & $ 15.6 $ & $ 0.0284  $      \\[-2mm]
$ 5 $ & $ 1 $ & $ 12.7 $ & $ 13.0 $ & $ 0.0593  $     &
$ 5 $ & $ 4 $ & $ 21.2 $ & $ 19.4 $ & $ 0.0332  $      \\[-2mm]
$ 6 $ & $ 1 $ & $ 14.7 $ & $ 15.2 $ & $ 0.0560  $     &
$ 6 $ & $ 4 $ & $ 23.7 $ & $ 22.2 $ & $ 0.0359  $      \\[-2mm]
$ 7 $ & $ 1 $ & $ 16.4 $ & $ 17.0 $ & $ 0.0540  $     &
$ 7 $ & $ 4 $ & $ 26.7 $ & $ 25.3 $ & $ 0.0337  $      \\[-2mm]
$ 8 $ & $ 1 $ & $ 17.9 $ & $ 18.5 $ & $ 0.0510  $     &
$ 8 $ & $ 4 $ & $ 28.6 $ & $ 27.4 $ & $ 0.0337  $      \\[-2mm]
$ 9 $ & $ 1 $ & $ 19.3 $ & $ 19.8 $ & $ 0.0496  $     &
$ 9 $ & $ 4 $ & $ 29.7 $ & $ 28.5 $ & $ 0.0344  $      \\[-2mm]
$10 $ & $ 1 $ & $ 20.2 $ & $ 20.7 $ & $ 0.0474  $     &
$10 $ & $ 4 $ & $ 31.1 $ & $ 30.0 $ & $ 0.0329  $      \\[-2mm]
$12 $ & $ 1 $ & $ 20.5 $ & $ 21.0 $ & $ 0.0443  $     &
$12 $ & $ 4 $ & $ 33.8 $ & $ 32.9 $ & $ 0.0305  $      \\[1mm]

$ 4 $ & $ 6 $ & $ 18.6 $ & $ 16.0 $ & $ 0.0275  $      &
$ 4 $ & $ 8 $ & $ 18.8 $ & $ 15.7 $ & $ 0.0273  $      \\[-2mm]
$ 5 $ & $ 6 $ & $ 21.9 $ & $ 19.6 $ & $ 0.0323  $      &
$ 5 $ & $ 8 $ & $ 22.1 $ & $ 19.3 $ & $ 0.0330  $      \\[-2mm]
$ 6 $ & $ 6 $ & $ 24.6 $ & $ 22.6 $ & $ 0.0340  $      &
$ 6 $ & $ 8 $ & $ 25.0 $ & $ 22.5 $ & $ 0.0340  $      \\[-2mm]
$ 7 $ & $ 6 $ & $ 27.5 $ & $ 25.5 $ & $ 0.0333  $      &
$ 7 $ & $ 8 $ & $ 27.7 $ & $ 25.4 $ & $ 0.0339  $      \\[-2mm]
$ 8 $ & $ 6 $ & $ 29.5 $ & $ 27.6 $ & $ 0.0339  $      &
$ 8 $ & $ 8 $ & $ 29.5 $ & $ 27.4 $ & $ 0.0325  $      \\[-2mm]
$ 9 $ & $ 6 $ & $ 30.9 $ & $ 28.2 $ & $ 0.0327  $      &
$ 9 $ & $ 8 $ & $ 31.0 $ & $ 28.9 $ & $ 0.0324  $      \\[-2mm]
$10 $ & $ 6 $ & $ 32.4 $ & $ 30.8 $ & $ 0.0320  $      &
$10 $ & $ 8 $ & $ 32.9 $ & $ 30.8 $ & $ 0.0321  $      \\[-2mm]
$12 $ & $ 6 $ & $ 34.2 $ & $ 32.7 $ & $ 0.0317  $      &
$12 $ & $ 8 $ & $ 34.5 $ & $ 32.7 $ & $ 0.0318  $      \\
   
\end{tabular}
\end{center}
\end{table}

\begin{figure}
\vspace{-4cm}
\centering{\includegraphics[width=12cm,clip=]{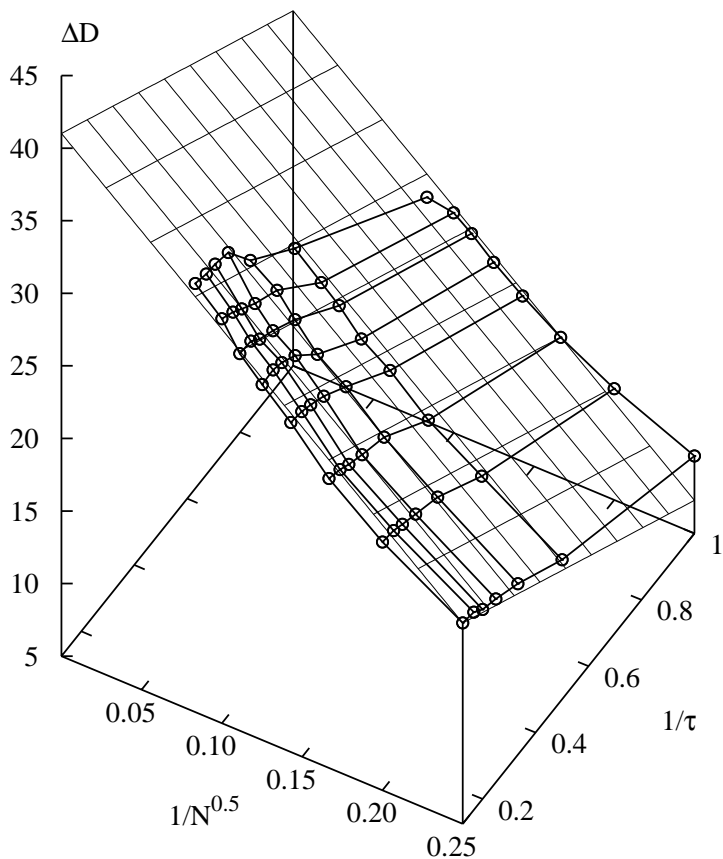}}
\vspace{-1cm}
\caption{
Comparison of the fit $ \Delta D = 42.7 - (89/N^{1/2}) - (13.2/\tau )$
to the simulation results for the dimensionality loss in the
full $4N+2$-dimensional phase space. 
}
\label{fig-3}
\end{figure}

	Our own previous work, on color conductivity and shear flow \cite{b11},
strongly suggests deviations in the Lyapunov spectrum of order the
inverse system width, $\sqrt{1/N}$ in two dimensions.  The present results
are roughly consistent with this finding though an even slower variation
with $N$ provides a comparable fit.  We know of no previous systematic
study of the variation of the Lyapunov spectrum with $\tau $.  A cursory
investigation shows that the dimensionality loss varies roughly as
$\tau ^{-1}$.  Thus the dimensionality loss, with fixed boundary temperatures
of 0.009 and 0.001, can be represented by 
$$ 
  \Delta D = 42.7(5) - \frac{89(2)}{\sqrt{N}} - \frac{13.2(5)}{\tau}\;. 
$$
The standard deviation affecting the last digits of the fit parameters
are given in brackets.
We compare the second expression with our numerical data in Fig. \ref{fig-3}.  
The rather good fit
indicates that the dimensionality loss persists in the large-system limit.
We investigate the loss further in the following section.

\section{Hamiltonian Projection: $\Delta D_{\mathcal{H}}$}

	Ten years ago \cite{b14} we studied the {\em rotation} rate of the $n$
phase-space
offset vectors $\{ \delta _j\} $.  We found that the rotation rate increases
very rapidly with system size, soon becoming very large relative to the
Lyapunov exponents themselves.  This observation suggests that the instantaneous
growth and decay rates in phase space---the instantaneous Lyapunov
exponents---might become isotropic in the large-system limit.  This suggests
that the measured growth and decay rates are also closely associated with the
subspaces spanned by the corresponding phase-space offset vectors
$\{ \delta\}$.  Because the instantaneous Lyapunov exponents measure
{\em radial} expansion and contraction, without any explicit rotational
contribution, the instantaneous growth rates
$\{ \lambda _j(t) = \Lambda _{jj}  \} $
associated with every one of the phase-space directions contributing to a
particular vector are identical.  The contributions in a
fixed phase-space direction are the summed-up contributions from the entire
set of Lyapunov vectors.

	At a particular phase-space point these contributions of the 
principal axes of the dynamical matrix
to the Lyapunov exponents are all proportional to $\cos ^2(\theta )$, where
$\theta $ is the angular difference between the principal axis and the
corresponding instantaneous $\delta $ vector.  
Because the Lyapunov exponents measure
growth or decay in the direction of $\delta $, the same logarithmic
growth rate, $(d/dt)\ln \delta $, applies to each component of the vector.

	In the {\em full} phase space the information dimension of the strange
attractor is given by the number of Lyapunov exponents whose sum is zero:
$$ \sum _{j=1}^{k}\lambda _j = 0 \longrightarrow D = k\ ;\ \Delta D = 4N+2-k\ .$$
It has to be emphasized that $k$ (as well as $k_{\mathcal{H}}$ introduced below) is
{\em not} an integer.  A linear interpolation between two successive values of
the Lyapunov sum is implied, with $k$ chosen such that the interpolated sum is
precisely zero.
In the Hamiltonian subspace the Lyapunov exponents contribute according to
their projections into that space:
$$ \sum _{j=1}^{k_{\mathcal{H}}}\langle \cos ^2(\theta _j)\lambda _j(t)
 \rangle = 0
\longrightarrow
D_{\mathcal{H}} = \sum _{j=1}^{k_{\mathcal{H}}}\langle \cos ^2(\theta _j)
  \rangle  \ ;$$
$$\Delta D_{\mathcal{H}} =
\sum _{j=k_{\mathcal{H}}}^{4N+2}\langle \cos ^2(\theta _j)\rangle =
4N-8-\sum_{j=1}^{k_{\mathcal{H}}}\langle \cos ^2(\theta _j)\rangle \ .$$
The $\cos ^2(\theta )$ form of the projection is required by the condition that
changing the sign of the offset vectors leaves the growth rate unchanged.  This
form follows from the quadratic form describing an infinitesimal phase-space
hyperellipsoid centered on a moving trajectory point.  Note
also that this weighting satisfies the normalization of the projections:
$$\sum _{j=1}^{4N+2}\cos ^2(\theta_j) \equiv 4N - 8\ .$$
Partial sums give an effective number of exponents in the Hamiltonian projection
of the full phase space.  Thus the analog of Kaplan and Yorke's conjecture
for the Hamiltonian subspace is the effective number of exponents,
$\sum \langle \cos ^2(\theta )\rangle $,
at which the projected sum, $\sum \langle \lambda (t)\cos ^2(\theta )\rangle $,
vanishes. 

	We explored the Kaplan-Yorke analog for the present problem, computing
the delta vectors and their projections.  The results are interesting.  First,
we noted that the summed-up local Lyapunov exponents are not strongly correlated
with the directions in the subspace:
$$\sum \lambda \langle \cos ^2(\theta )\rangle \simeq
\sum \langle \lambda (t)\cos ^2(\theta )\rangle \ .$$
This suggests that all the time-averaged projections of the various vectors into
the Hamiltonian subspace are similar,
$$\langle \cos ^2(\theta )\rangle \simeq (4N-8)/(4N+2)\ .$$
See Figs. 4 and 5.
\begin{figure}
\centering{\includegraphics[width=12cm,clip=]{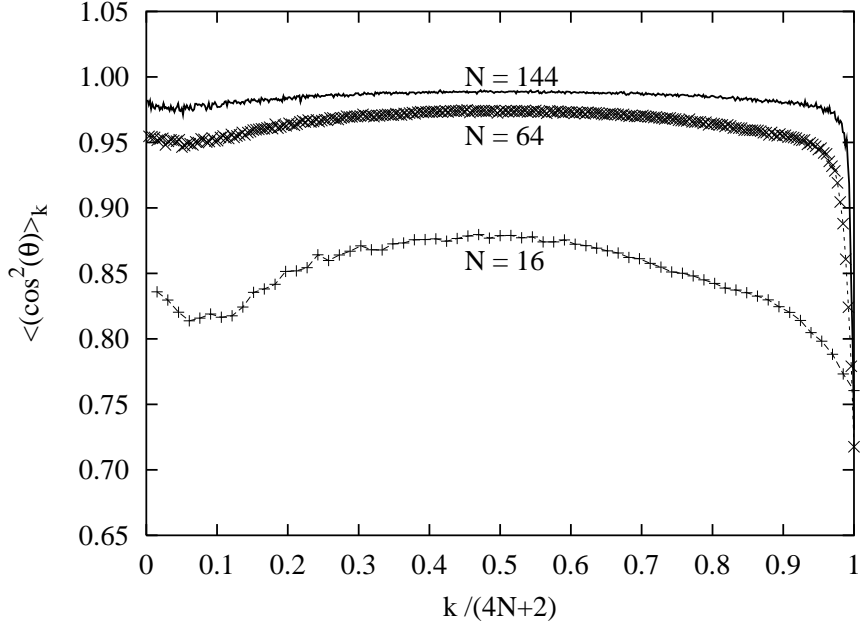}}
\caption{
Time-averaged projections $\{ \langle \cos ^2(\theta )\rangle _k\} $
of the full-space $\{ \delta _k\} $ into the Hamiltonian portion of
phase space for $4\times 4$, $8\times 8$, and $12\times 12$ particles.
The Nos\'e-Hoover time $\tau$ is $6.$
It is evident that in the large-system limit the influence of the boundary
degrees of freedom disappears.
}
\label{fig-4}
\end{figure}
\begin{figure}
\centering{\includegraphics[width=12cm,clip=]{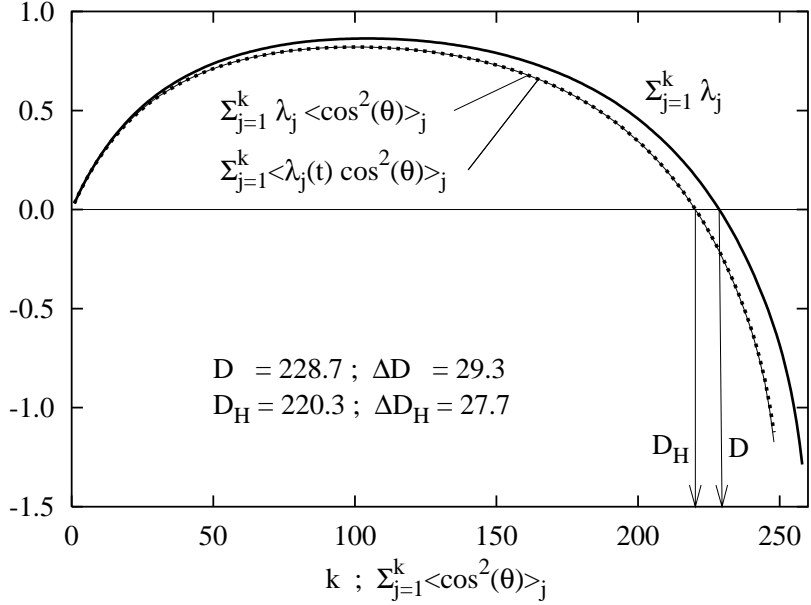}}
\caption{
Typical summed Lyapunov spectra for the full space and for two estimates of
the Hamiltonian projection.  The data shown here, for 64 particles and for
a thermostat response time $\tau = 5$,
show partial sums of the 258 Lyapunov exponents, $\sum _{j \le k}\lambda _j$,
and the two projections,
$ \sum _{j\le k}\lambda _j\langle \cos ^2(\theta _j)\rangle \ ;\
  \sum _{j\le k}\langle \lambda_j(t) \cos ^2(\theta _j)\rangle \ , $
as functions
of $k$ and $\sum _{j \le k}\langle \cos ^2(\theta _j)\rangle $.  
The two projections are actually different but indistinguishable on this scale. 
Kaplan and
Yorke's form for the information dimension is the linearly-interpolated
value of $k$ for which the linearly-interpolated sum vanishes.  The analog,
for the Hamiltonian portion of phase space, is the interpolated value of
$\sum \cos ^2(\theta )$ at which the corresponding projected sum of exponents
vanishes.
}
\label{fig-5}
\end{figure}
Fig. \ref{fig-4} demonstrates that the projection of the vectors 
becomes increasingly
uniform as system size is increased, and is quite close to the average value,
$(4N - 8)/(4N + 2)$ expected for fully random projection directions.  Fig. 5
compares the two estimates for the projected Lyapunov sums,
$$\sum \lambda \langle \cos ^2 (\theta )\rangle \ {\rm and }\
  \sum \langle \lambda (t)                \cos ^2 (\theta )\rangle \ .$$
The two sums vanish at nearly the same projected dimensionality,
$\sum \cos ^2(\theta )$,
indicating that the correlation of the exponents with direction is small.

	We have used the analog of the Kaplan-Yorke formula to estimate the
dimensionality reduction in the Hamiltonian subspace, $\Delta D_{\mathcal{H}}$,
and show these results in Fig. 6.  Just as in the full phase space, the loss
\begin{figure}
\vspace{-3cm}
\centering{\includegraphics[width=12cm,clip=]{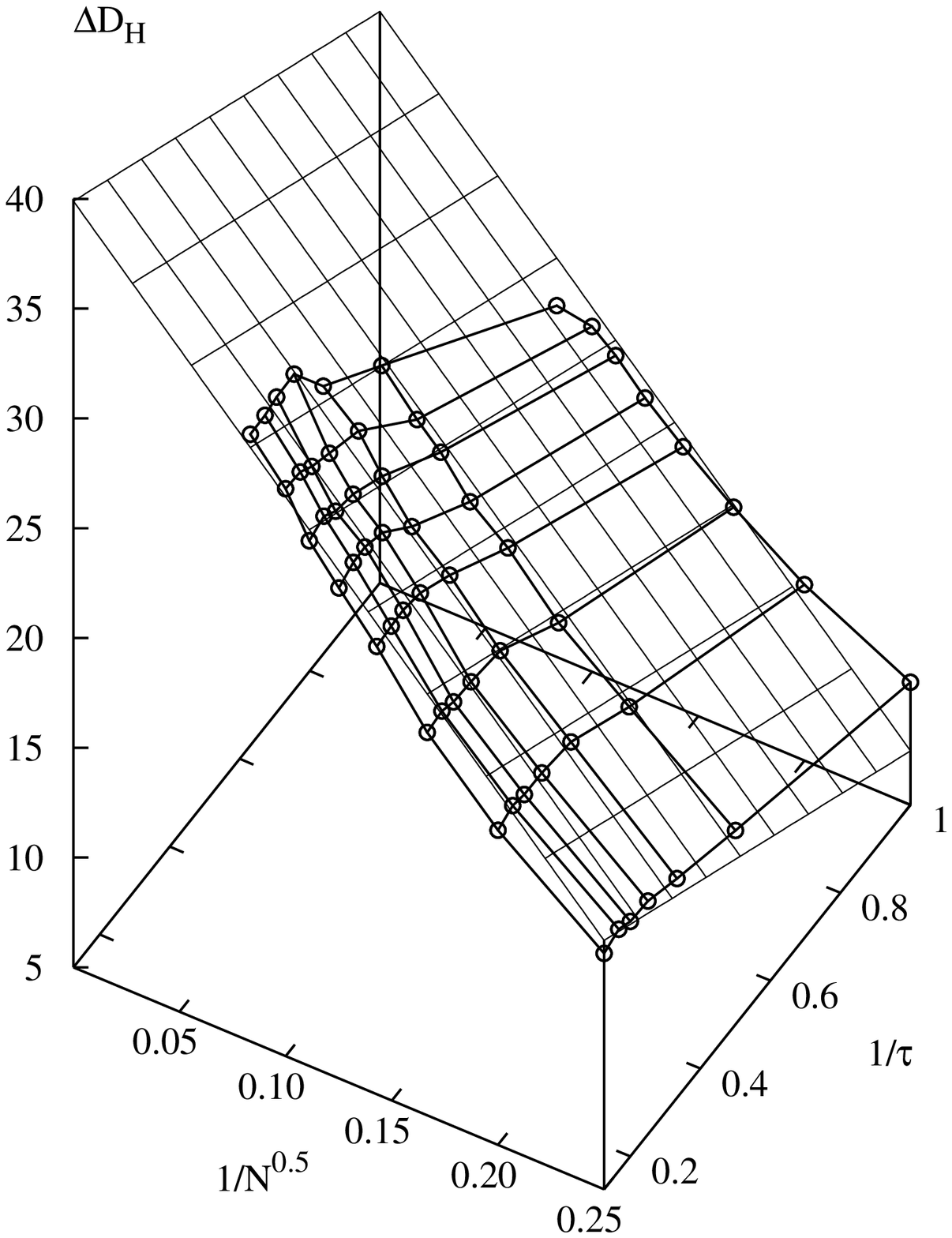}}
\vspace{-1cm}
\caption{
Comparison of the dimensionality loss $\Delta D_{\mathcal{H}}$ in the 
Hamiltonian subspace to a fit,
$ \Delta D_{\mathcal{H}} = 41.2 - 94 /\sqrt{N} - 10.1 /\tau \; , $ to the data.
}
\label{fig-6}
\end{figure}
of dimensionality varies smoothly with relaxation time and system size,
and may be represented by
$$ 
  \Delta D_{\mathcal{H}} = 41.2(5) - \frac{94(3)}{\sqrt{N}} - \frac{10.1(5)}
      {\tau}\;. 
$$

	It is worth pointing out that a na\"{\i}ve approach to dimensionality
loss in the Hamiltonian subspace could be based on an orthonormalization of
the Hamiltonian subspace only, propagating $\delta $ vectors with the 
Hamiltonian
equations of motion but using the thermostated equations of motion for the
underlying reference trajectory.  This approach, which we explored years ago
and which has recently been considered independently by Ken Aoki [private
communications, 2002], is equivalent
to considering the Lyapunov spectrum for a Hamiltonian system subject to
time-dependent forces $\{ F(t)\} $, where the forces are due to the thermostated
particles.  Such equations of motion, being Hamiltonian, satisfy Liouville's
Theorem \cite{b9,b18}, and guarantee that the corresponding Lyapunov spectrum is
made up of pairs of positive and negative exponents, with zero sum.

\section{Conclusions}

	The ``$\phi ^4$" model shows conclusively that the nonequilibrium
steady-state loss of phase-space dimensionality can easily exceed the
dimensionality associated with the system boundaries.  The present
results also confirm that the deviations from the large-system limit
vary according to a power law, $N^{-p}$, $1/4 \le p \le 1/2$,
roughly compatible with the inverse of the system size, 
$1/\sqrt{N}$ in two dimensions. 
For $\tau \le 5$  they
have in addition deviations in the relaxation frequency
$\propto (1/\tau )$.  The quartic tethers are a particularly
useful feature of the model, which make it possible to carry out the
simulations without the need to take thermal expansion explicitly into
account.

     The dimensionality loss $\Delta D$ is expected to be
extensive as has been stressed in the introduction. For
fixed thermostat temperatures $T_H$ and $T_C$ the temperature
gradient is determined by the system size, proportional to $1/\sqrt{N}$ 
in two dimensions, and 
$\Delta D$ becomes independent of $N$. This is indeed the case 
in the large-particle limit  $N \to \infty$. Our simulations thus
confirm the extensivity of the dimensionality reduction for
stationary heat flow in the linear-response limit. Far from
equilibrium weak deviations from this $N$-dependence are found.

        Table 1 reveals that for short Nos\'e-Hoover relaxation times
the dimensionality loss 
$\Delta D_{\mathcal{H}}$ in the purely Hamiltonian part of phase space
may even slightly exceed the dimensionality loss $\Delta D$ in the full phase 
space. This is because  $\Delta D_{\mathcal H}$  is overestimated by one
due to the lack of energy conservation in the Hamiltonian subspace
once the thermostats are added.  

	The projection of the phase-space offset vectors $\{ \delta\} $
into the Hamiltonian subspace developed here shows that most of the
dimensionality loss occurs in a part of the system which obeys purely
conservative Hamiltonian equations of motion.  Rapid rotation is
responsible.  This rotation nearly eliminates the correlations between
phase-space contraction and direction.  Evidently phase-space contraction
is not only real, but relatively simple, and certainly must persist in
the large-system thermodynamic limit.  Thus the present results corroborate
our interpretation of the second law of thermodynamics for nonequilibrium
stationary states \cite{b8}.  Such states occupy not just a reduced volume
in phase space.  They are restricted to a subspace of reduced dimensionality,
with the dimensionality loss simply related to the rate of external
thermodynamic entropy production.  Useful models illustrating
dimensionality loss for periodic color conducting or shear flows can now
be developed as extensions of this idea.

\section*{Acknowledgements}

         Work at the University of Vienna was supported by the Austrian
Fonds zur F\"orderung der wissenschaftlichen Forschung, Project P-15348.
Work at Livermore was performed under the auspices of the United States
Department of Energy through University of California Contract W-7405-Eng-48.
We very much appreciate the interest and encouragement of Kenichiro Aoki,
Aurel Bulgac, Carol Hoover, Dimitri Kusnezov, and David Ruelle.


w
\end{document}